\documentclass[fleqn,usenatbib]{mnras}

\usepackage[T1]{fontenc}

\DeclareRobustCommand{\VAN}[3]{#2}
\let\VANthebibliography\thebibliography
\def\thebibliography{\DeclareRobustCommand{\VAN}[3]{##3}\VANthebibliography}

\usepackage{graphicx}	
\usepackage{amsmath}	
\usepackage{amssymb}	

\def \aj {AJ}
\def \mnras {MNRAS}
\def \apj {ApJ}

\def \apjl {ApJL}
\def \aap {A\&A}

\def \araa {ARAA}
\def \pasp {PASP}

\usepackage{newtxtext,newtxmath}

\title[NOEMA mm detection of a NS HMXB]{The first mm detection of a neutron star high-mass X-ray binary}

\author[J. van den Eijnden et al.]{
J. van den Eijnden,$^{1}$
L. Sidoli,$^{2}$
M. D\'iaz Trigo,$^{3}$
N. Degenaar,$^{4}$
I. El Mellah,$^{5,6}$
F. F\"urst,$^{7}$
V. Grinberg,$^{8}$
\newauthor P. Kretschmar,$^{9}$
S. Mart\'inez-N\'u\~nez,$^{10}$
J. C. A. Miller-Jones,$^{11}$
K. Postnov,$^{12,13}$
T. D. Russell,$^{14}$
\\
$^{1}$Department of Physics, University of Warwick, Coventry CV4 7AL, UK\\
$^{2}$INAF, Istituto di Astrofisica Spaziale e Fisica Cosmica, Via A. Corti 12, 20133, Milano, Italy\\
$^{3}$ESO, Karl-Schwarzschild-Strasse 2, 85748, Garching bei M\"unchen, Germany\\
$^{4}$ Anton Pannekoek Institute for Astronomy, University of
Amsterdam, Science Park 904, Amsterdam, 1098 XH, The
Netherlands\\
$^{5}$Departamento de F\'isica, Universidad de Santiago de Chile, Av. Victor Jara 3659, Santiago, Chile \\
$^{6}$ Center for Interdisciplinary Research in Astrophysics and Space Exploration (CIRAS), USACH, Chile\\
$^{7}$ Quasar Science Resources SL for ESA, European Space Astronomy Centre (ESAC), Science Operations Departement, 28692, Villanueva de la Cañada, Madrid, Spain\\
$^{8}$ European Space Agency (ESA), European Space Research and Technology Centre (ESTEC), Keplerlaan 1, 2201 AZ, Noordwijk, The Netherlands\\
$^{9}$ European Space Agency (ESA), European Space Astronomy Centre (ESAC), Camino Bajo del Castillo s/n, 28692, Villanueva de la Cañada, Madrid, Spain\\
$^{10}$ Instituto de Física de Cantabria (CSIC-Universidad de Cantabria), 39005, Santander, Spain\\
$^{11}$ International Centre for Radio Astronomy Research, Curtin University, GPO Box U1987, Perth, WA 6845, Australia\\
$^{12}$ M.V. Lomonosov Moscow State University, Sternberg Astronomical Institute, 13, Universitetskij pr., 119234, Moscow, Russia\\
$^{13}$ Kazan Federal University, Kremlevskaya 18, 420008 Kazan, Russia\\
$^{14}$ Istituto di Astrofisica Spaziale e Fisica Cosmica, INAF, Via U.
La Malfa 153, Palermo, I-90146, Italy\\
}

\date{Accepted XXX. Received YYY; in original form ZZZ}

\pubyear{2021}

\begin{document}
\label{firstpage}
\pagerange{\pageref{firstpage}--\pageref{lastpage}}
\maketitle

\begin{abstract}
Neutron stars accreting from OB supergiants are often divided between persistently and transiently accreting systems, called Supergiant X-ray Binaries (SgXBs) and Supergiant Fast X-ray Transients (SFXTs). This dichotomy in accretion behaviour is typically attributed to systematic differences in the massive stellar wind, binary orbit, or magnetic field configuration, but direct observational evidence for these hypotheses remains sparse. To investigate their stellar winds, we present the results of pilot 100-GHz observations of one SFXT and one SgXB with the Northern Extended Millimetre Array. The SFXT, IGR J18410-0535, is detected as a point source at $63.4 \pm 9.6$ $\mu$Jy, while the SgXB, IGR J18410-0535 remains undetected. Radio observations of IGR J18410-0535 imply a flat or inverted low-frequency spectrum, arguing for wind emission and against non-thermal flaring. Due to the uncertain SFXT distance, however, the observations do not necessarily imply a difference between the wind properties of the SFXT and SgXB. We compare the mm constraints with other HMXBs and isolated OB supergiants, before considering how future mm campaigns can constrain HMXB wind properties by including X-ray measurements. Specifically, we discuss caveats and future steps to successfully measure wind mass loss rates and velocities in HMXBs with coordinated mm, radio, and X-ray campaigns.
\end{abstract}

\begin{keywords}
accretion: accretion disks -- stars: individual (IGR J18410-0535, AX J1841.0–0536; X1908+075, 4U 1909+07) -- stars: neutron -- X-rays: binaries -- stars: massive -- stars: mass-loss
\end{keywords}

\section{Introduction}

Massive stars and their winds play a central role in a wide range of astrophysical contexts. The mass-loss effects of these stellar winds can significantly impact their evolution, supernova properties, and resulting compact object type and properties \citep[][]{eldridge2004,belczynski2010}. Impacting their surroundings, massive star feedback shapes the stellar clusters that host them \citep{Prajapati2019}, for instance driving large-scale wind-blown bubbles and shocks that accelerate particles and possibly Galactic cosmic rays \citep{Aharonian2019}, as well as impacting the overall evolution of their host galaxy. Stellar winds may also attenuate signals passing through, for instance from orbiting pulsars, through depolarization and dispersion \citep{Wang2022,Li2022}.

In binary systems, stellar winds affect mass transfer and accretion, furthermore impacting binary evolution and the formation rates of gravitational wave merger precursor systems \citep{vandenheuvel2017}. These effects are particularly visible in binary systems called high-mass X-ray binaries (HMXBs): a compact object orbiting closely around a massive star and persistently or transiently accreting mass from its companion. A relatively short-lived binary evolutionary phase, HMXBs can be classified in various types. Based on the type of donor star, systems with OB supergiant or Be star donors are typically separated. The latter type, called Be/X-ray binaries, are predominantly transient \citep{reig2011}, while the former may be either transient of persistent. The transients with a supergiant donor, referred to as supergiant fast X-ray transients, or SFXTs, reside in weakly accreting states with $L_X \lesssim 10^{34}$ erg/s for the majority of time \citep[$\gtrsim 95-99$\%;][]{Sidoli2018}, interspersed by brief X-ray flares reaching $L_X \sim 10^{35}-10^{37}$ erg/s. On the other hand, in persistent systems with a supergiant donor, supergiant X-ray binaries (SgXBs), the compact object accretes persistently from the OB star's wind at typical levels of $L_X \sim 10^{35}-10^{36}$ erg/s. Regardless of HMXB class, their compact objects are predominantly identified as neutron stars with strong, $B>10^{12}$ G magnetic fields and slow, $P > 1$ second spins, with only a handful of exceptions \citep[see e.g.,][for recent catalogues]{neumann2023,fortin2023}. 

Despite their similar donor star types, SFXTs and SgXBs evidently display distinct accretion behaviour. The origin of this difference remains an open question, residing in the interaction of the gravitationally captured stellar wind with the neutron star magnetosphere, where the wind properties and/or the binary orbital geometry play a crucial role (see \citealt{Kretschmar2019} for a recent review).

The winds of massive stars are line-driven through resonant absorption of the star’s UV photons by metals in the outer envelope of the star. The winds accelerate away from the star, thereby tapping into higher energy UV photons due to Doppler shifts, until they reach their terminal velocity $v_{\infty}$ at a distance of a few stellar radii \citep{lucy1970}. Together with the mass loss rate $\dot{M}_{\rm wind}$, this terminal velocity forms the principal global property of the wind \citep{puls2008}. The winds are not smooth, but instead show micro-structure: on small scales, instabilities that are inseparably connected to the line-driving process lead to shocks and the formation of overdense regions, known as clumps. These three wind properties -- mass loss rate, velocity, and clumpiness -- in combination with the size and eccentricity of the orbit may give rise to different accretion regimes. The clumps, often assumed to be responsible for the X-ray flares in SFXTs \citep{bozzo2016,Sidoli2017}, are also known to be present in the stellar winds in SgXBs \citep[e.g.][]{torrejon2015,grinberg2017,armato2021,diez2023}. Furthermore, numerical simulations of clumpy wind accretion have shown that the bow shock formed around the compact object smears out the inhomogeneities, which questions whether the X-ray flares can really be ascribed to the serendipitous capture of a clump \citep{elmellah2018}. 

Observational characterizations of the stellar winds in neutron star HMXBs, as a means to test the hypothesis of systematic differences between SFXT and SgXB winds, remain relatively rare to date. Comparing the prototypical SgXB Vela X-1 with the prototypical SFXT IGR J17544-2619 using IR, optical, and UV observations, \citet{gimenez2016} reported significant differences between the inferred wind velocity in the two systems (700 versus 1500 km s$^{-1}$, respectively). However, \citet{hainich2020} found no evidence in optical and UV spectra for such a dichotomy in the wind properties of a slightly larger sample of HMXBs. Instead, differences in the orbital period and eccentricity were suggested to play a significant role in driving the accretion mode. Beyond remaining inconclusive across the literature, these comparisons have been limited to methods that may suffer from systematic uncertainties. Wind diagnostic methods exist across a wide range of wavelengths, from the radio to the UV band and even in X-rays \citep[e.g.,][]{elmellah2020}. However, wind clumping may lead to mis-estimates of global wind properties by an order of magnitude \citep{fullerton2006}, particularly affecting IR/optical/UV/X-ray studies: the radial distance in the stellar wind scales with observed wavelength, while wind clumping decreases with distance, leaving such studies particularly affected.

Arising from the outer, (mostly) unclumped wind regions, the thermal radio and mm wind emission may be used instead as a different estimator of the global wind properties \citep{lamers1998,puls2006,martinez2017}. Such low-frequency observations of isolated (nearby) OB supergiants have been used to infer their wind properties for several decades \citep{leitherer1991,gudel2002,fenech2018}; more recently, neutron star HMXBs have been detected at radio frequencies for the first time too \citep[see e.g.,][]{vandeneijnden2021}. Compared to the isolated case, the low-frequency range may include additional emission processes driven by accretion (e.g., radio jets). Due to the strongly-inverted spectrum of thermal stellar wind emission \cite[$\alpha = 0.6$, where $S_\nu \propto \nu^\alpha$;][]{wright1975}, the mm band of HMXBs may offer a view of their stellar winds. However, as their typical distances significantly exceed those of close-by isolated OB supergiants, successful mm studies of neutron star HMXBs have not been executed. Here, we present a pilot mm study of two neutron star HMXBs to assess whether this band can feasibly be used to detect and study their stellar winds. We show the results of deep, pilot Northern Extended Millimetre Array (NOEMA) mm observations of one SgXB, X1908+075, and one SFXT, IGR J18410-0535, complemented with radio observations for the latter. In this letter, we present the first mm detection of a SFXT, compare it with isolated massive stars and other HMXBs, and assess what future steps are necessary to succesfully use the mm band as a HMXB wind probe.

\section{Targets and observations}

\subsection{Targets}

X1908+075 (also 4U 1909+07) is a persistent SgXB hosting an X-ray pulsar ($\sim$604 s, \citealt{levine2004, jaisawal2020}) and a B0-B3 supergiant companion \citep{martinez2015}.
The system has an orbital period of 4.4 days an eccentricity of 0.021$\pm{0.036}$ \citep{wen2000, levine2004}.
The distances reported in Gaia eDR3 \citep{Bailer-Jones2021} for the optical counterpart (Gaia ID 4306419980916246656) differ between the geometric distance based on the parallax (d$_{\rm geo}$=5.0$^{+3.8} _{-2.1}$ kpc), and the photogeometric one (d$_{\rm pgeo}$=8.0$^{+3.0 } _{-4.0}$ kpc) determined by including the stellar photometry. Given these large uncertainties, we instead adopt here the distance of 4.85$\pm{0.50}$ kpc determined by \citet{martinez2015}, derived independently using photometric methods resulting in a consistent value with smaller uncertainties. 

IGR J18410-0535 (also AX J1841.0-0536; \citealt{Bamba2001}) 
is a SFXT \citep{Bozzo2011, romano2011, Sidoli2018} 
associated with a B-type supergiant donor (2MASS 18410043-0535465; \citealt{Halpern2004b}). The spectral classification reported in the literature differs slightly, with different estimated distances as a result: a B1Ib star at 3.2$^{+2.0} _{-1.5}$ kpc \citep{nespoli2008}; a B0.2 Ibp star at about 4 kpc \citep{Negueruela2008}; a B1I star at 6.9$\pm{1.7}$ kpc \citet{Sguera2009}. 
We note that the geometric and photogeometric distances listed in Gaia eDR3 (Gaia ID 4256500538116700160; \citealt{Bailer-Jones2021}) are very similar to each other (d$_{\rm geo}$=13.9$^{+3.7} _{-2.7}$ kpc  and  d$_{\rm pgeo}$=13.7$^{+4.0} _{-2.8}$ kpc), but they are much larger than non-parallax values. A pulsar periodicity originally proposed by \citet{Bamba2001} has been posed into question by \citet{Bozzo2011}. A tentative orbital period of 6.45 days with $e=0.16\pm0.11$ awaits a confirmation \citep{Gonzalez2014}.

\subsection{mm-band: NOEMA}

We observed X1908+075 and IGR J18410-0535 with NOEMA across multiple observing runs as part of observing program W22BO. X1908+075 was observed twice (W22BO-001), on January 25th and February 7th, 2023, for a total on-source observing time (i.e. excluding overheads and calibration) of 1.46 hours yielding an $11.4$ $\mu$Jy/bm sensitivity. IGR J18410-0535 was observed across three runs (W22BO-002), on February 20th, 22nd, and 23rd, 2023, for the same total on-source time of 1.46 hours resulting in a $9.6$ $\mu$Jy/bm sensitivity. Both sources were observed with a continuum-only point-source-detection setup at 100 GHz, using a lower sub band (LSB) sensitive between 82.5 and 90 GHz and an upper sub band (USB) sensitive between 98 and 105.5 GHz. A standard calibration setup was used, with initial flux and bandpass calibrator scans, followed by phase calibrator scans book-ending target scans. 

We then reduced the data using the standard suite of software for NOEMA data, \textsc{gildas}, using a combination of the standard \textsc{continuum and line interferometer calibration} (\textsc{clic}) pipeline and manual inspection and data quality assessments. Each observing run was calibrated individually, after which we combined the resulting calibrated data into one uv-table per target. Due to the low signal-to-noise of our targets (see below), the individual runs were not sufficiently sensitive for further analysis. The analysis of the resulting combined, calibrated uv-tables, including frequency-averaging and merging both sub bands, was performed in \textsc{mapping} within \textsc{gildas}. Specifically, we used the \textsc{mapping} task \textsc{uv\_fit} to attempt to fit a point source in each sub band at a location consistent with the pointing centre within one synthesized beam. We then merged the bands and repeated the fitting attempt for the full, merged band, to either improve the signal to noise of a detection in the sub bands or to search for a detection at higher sensitivity if the sub bands did not reveal a source. Finally, the RMS sensitivity of each data set was measured using \textsc{mapping}. 

\subsection{Radio band: ATCA}

At cm wavelengths, we performed DDT observations of the field of IGR J18410-0535 with the Australia Telescope Compact Array (ATCA) on February 25th, 2023, i.e. 2 days after the last NOEMA run. Due to a lack of suitable ATCA time available for DDT requests around the NOEMA runs on X1908+075, we did not observe this target at radio wavelengths. The observations of IGR J18410-0535 were carried out using six equal-length target scans during an observing run for an unrelated project (C3493; PI Van den Eijnden; arrray configuration: 1.5B), spread out across the run to maximize uv-plane coverage despite the low on-source time. The total on-source observing time added up to 1.5 hours. As bandpass and phase calibrators, we used the standard calibrator PKS B1934-638 and B1829-106, respectively. Data were recorded at 5.5 and 9 GHz simultaneously with 2 GHz of bandwidth at each frequency. We imported the native ATCA data files into the \textsc{common astronomy software application} \citep[\textsc{casa};][]{casa2022} v6.5.5.21, after which we followed standard procedures to flag, calibrate, and image the data. We imaged the data using the \textsc{tclean} task in \textsc{casa} using a Briggs weighting scheme and a robust parameter of one. As no significant emission was detected at either frequency band, we produced a single deep image combining both bands. Still, no counterpart was detected. The RMS sensitivity, calculated across a region covering the source position and devoid of other point sources, was $16$ $\mu$Jy/bm. 

\subsection{X-ray band: \textit{MAXI}}

To assess the level of accretion-induced X-ray luminosity and the presence X-ray flaring, we also investigated the X-ray monitoring data from the \textit{Monitor of All-sky X-ray Image} \citep[\textit{MAXI;}][]{matsuoka2009}/Gas Slit Camera (GSC) aboard the International Space Station. For both targets, we used the on-demand tool on the instrument's dedicated webpage\footnote{\url{http://maxi.riken.jp/top/index.html}} to extract a light curve and spectrum. We confirmed that no confusing point sources affect the source or background extraction regions of either source, despite \textit{MAXI}'s relatively limited spatial resolution. However, due to the location of the SFXT IGR J18410-0535 arcminutes away from the Galactic plane, its position is affected by an enhanced diffuse X-ray background of the plane. For both targets, we first extracted light curves at a daily cadence for the first six months of 2023, confirming that no significant levels of variability or flaring are seen at this cadence for either source. To check for the presence of shorter X-ray flares in the SFXT, we also extracted a light curve at a 6-hour cadence (balancing between typical flare time scales and detection sensitivity) for the two weeks centered at the dates of the NOEMA campaign. For the SgXB X1908+075, we combined all \textit{MAXI} from the first six months of 2023 to extract a spectrum. 

\section{Results}

\label{fig:lxlmm}


We detect significant 100-GHz emission from the SFXT IGR J18410-0535 at a level of $63.4 \pm 9.6$ $\mu$Jy, or a signal-to-noise of 6.6 (synthesized beam 2.67" $\times$ 0.35", position angle of 9.91 degrees). The position of the mm counterpart is consistent with the known target position, within the synthesized beam\footnote{The source peak position in the LSB data is consistent within its statistical fitting errors ($\sim10$ times smaller than the beam size) with zero offset from the Gaia position of source used as the pointing coordinates. The USB peak position is shifted slightly North by $0.6$", exceeding the position's statistical uncertainty but still well within the beam ($2.7$" in that direction). Such shifts can arise from phase calibration uncertainties, as NOEMA calibration is performed per sub band. We therefore followed routine procedure and used \textsc{uv\_shift} in \textsc{mapping} to shift the phase center in the USB to the peak position, before merging sub bands. This approach prevents the minor calibration shift from effectively decreasing the source flux density in the merged band.}. For a source distance of 3.2 kpc \citep{nespoli2008}, this detection implies a 100-GHz luminosity of $L_{\rm mm} = (7.8 \pm 1.2)\times10^{28}$ erg/s (but see Section \ref{sec:discussion} for the effect of larger distances). At the individual sub bands, we measure flux densities of $64.5 \pm 14.3$ (LSB) and $61.7 \pm 12.9$ $\mu$Jy (USB). In the radio band, the coordinated ATCA observations of IGR J18410-0535 did not reveal a counterpart, at a combined $5.5$+$9$ GHz band sensitivity of $16$ $\mu$Jy/bm, implying a 3$\sigma$ flux density upper limit of $48$ $\mu$Jy. Together, the ATCA and NOEMA measurements imply a 3-$\sigma$ lower limit on $\alpha$ (where $S_\nu \propto \nu^\alpha$) of $\alpha > -0.1$ \citep[following the method in Appendix B of][]{vandeneijnden2021}. Therefore, the mm and radio observations jointly imply a flat to inverted spectrum, and rule out the steep spectra ($\alpha \approx -0.5$) often associated with radio flaring in X-ray binaries \citep{fender2004outburststates}.

In contrast to IGR J18410-0535, we do not detect any significant mm emission from the targeted SgXB, X1908+075, in either of the sub bands or in the merged band. The combined band's RMS sensitivity is 11.4 $\mu$Jy/bm, implying a 3$\sigma$ upper limit of $34.4$ $\mu$Jy or $L_{\rm mm} < 9.7\times10^{28}$ erg/s assuming a source distance of 4.85 kpc \citep{martinez2015}. Due to the relatively short individual observing runs and low elevations, the uv-plane coverage for both targets is not sufficient to create a meaningful image, nor to analyse the individual runs.


\label{sec:results_xrays}

As a SgXB, X1908+075 has been systematically detected by a suite of X-ray instruments since its discovery. The systematic analysis of \textit{Rossi X-ray Timing Explorer} data by \citet{levine2004} reports that the source shows a variable flux along its orbit within a typical range of $(3-8)\times10^{-10}$ erg/s/cm$^{2}$ (2-30 keV). Later studies report consistent flux values \citep[e.g.,][]{furst2011,furst2012,jaisawal2020,Shtykovsky2022}. Such a flux level is faint but detectable with \textit{MAXI}; as the source's six-month daily light curve does not show significant variability, we fitted the target's spectrum accumulated across this time span. We then used \textsc{xspec} v12.13.0c \citep{arnaud1996} to fit the cumulative X-ray spectra and determine the X-ray flux. For our spectral fits, we assumed \citet{wilms2000} abundances and \citet{vern1996} cross sections. We applied $\chi^2$ statistics, as over the 6 months of combined monitoring (total on-source exposure of 176 ks), the target spectrum amassed $2.5\times10^4$ counts. However, due to the limited spectral resolution of \textit{MAXI} (40 bins between 2 and 20 keV), the spectral quality did not suffice to perform detailed spectral comparisons. Instead, we found that a simple absorbed power law model, \textsc{tbabs * power law}, provided a satisfactory fit, with a $\chi^2_\nu = 38.2/37$. Using the convolution model \textsc{cflux}, we measure an unabsorbed 1--10 keV flux of $(2.6\pm0.4)\times10^{-10}$ erg s$^{-1}$ cm$^{-2}$. We adopt this value, consistent with archival levels when accounting for the used energy bands, in the remainder of this work. At a distance of $4.85$ kpc, this flux implies $L_X = (7.3\pm1.1)\times10^{35}$ erg s$^{-1}$. 

The non-flaring X-ray level of IGR J18410-0535 falls several orders below the sensitivity of \textit{MAXI}. Its flaring levels, reaching fluxes of several times $10^{-10}$ erg s$^{-1}$ cm$^{-2}$ \citep[e.g.,][]{romano2011,Bozzo2011,nobukawa2012}, may reach up to $\sim 0.1$ ct s$^{-1}$ cm$^{-2}$ in the full 2-20 keV \textit{MAXI} band. However, in the six-hour cadence \textit{MAXI} light curve of the two weeks around the three NOEMA observations, no significant emission is observed. The six hour (i.e. $21.6$ ks) cadence matches the typical time scale of these brightest flares of IGR J18410-0535 \citep[e.g.,][]{romano2011} -- at shorter time scales, the loss of sensitivity would prevent the detection of flares further. Furthermore, based on fourteen years of \textit{INTEGRAL} monitoring data, \citet{Sidoli2018} estimate a $0.53$\% flaring duty cycle for IGR J18410-0535. Therefore, the probability of a flare occurring during each of the three NOEMA observing runs is negligible ($\sim10^{-7}$), and the probability of a flare having occured during the ATCA observation is low. 

\section{Discussion}
\label{sec:discussion}

\subsection{Comparisons to isolated OB supergiants and other HMXBs}

In this Letter, we have presented a pilot mm study of a SgXB and a SFXT. The main outcome of this campaign is the first ever detection of mm emission from a neutron star HMXB, in the SFXT IGR J18410-0535. While several SgXBs have been detected at radio frequencies, this result also constitutes the first detection of low-frequency (i.e., below IR) emission from a SFXT. A quasi-simultaneous radio observation of IGR J18410-0535 rules out that the mm emission arises from a low-frequency flare associated with an X-ray flare. Importantly, while the SgXB X1908+075 was not detected, the mm data does not necessarily imply a difference between the two targets or source classes: due to the large distance uncertainties of IGR J18410-0535, its luminosity can feasibly be lower than the measured upper limit for X1908+075.
 
Radio and (sub-)mm observations of isolated supergiants have long been used to investigate their stellar wind properties, which allows for a comparison with our HMXB results. As these studies are typically not performed at the same exact frequencies, we will compare specific luminosities: $(7.8 \pm 1.2)\times10^{17}(d/3.2\text{ kpc})^2$ and $<9.7\times10^{17}$ erg/s/Hz for IGR J18410-0535 and SgXB X1908+075, respectively. The ALMA 100-GHz study of the massive star population in Westerlund 1 by \citet{fenech2018}, detects sixteen OB supergiants in the range of $3.3\times10^{18}$--$3\times10^{19}$ erg/s/Hz, while \citet{leitherer1991} report the detection of seven OB supergiants at 230 GHz between $4\times10^{18}$--$10^{20}$ erg/s/Hz. While these mm comparisons may suggest our detection (depending on distance) and upper limit are deeper in specific luminosity than earlier studies, radio observations alter that conclusion. For instance, \citet{scuderi1998} detect thermal radio emission from eleven OB supergiants (all within 2 kpc) down to $3\times10^{17}$ erg/s/Hz, while three are not detected with upper limits down to $1.6\times10^{17}$ erg/s/Hz. Including the correction to the specific mm luminosity due to the strongly inverted wind spectrum (implying a factor $\sim 4.4$ in luminosity between $8.45$ and $100$ GHz), the detections and limits in \citet{scuderi1998} overlap in their range with our two targets.

While in isolated OB supergiants (but not in massive binaries), low-frequency emission is straightforwardly attributed to the stellar wind, we can briefly consider alternative origins in HMXBs as well. The low-frequency emission of HMXBs has been proposed to arise from thermal stellar wind emission, non-thermal accretion-driven jet emission, their combination, and their wind-jet interaction \citep[e.g.,][]{vandeneijnden2021,vandeneijnden2022,chatzis2022}. As the mm detection of IGR J18410-0535 was obtained in its quiescent state, a significant jet contribution is unlikely: for a flat spectrum jet, its inferred 6-GHz radio luminosity would be $\sim 5\times10^{27}$ erg/s, which approximately equals the radio jet luminosities of black hole X-ray binaries at similar (i.e. $10^{32}$--$10^{33}$ erg/s) quiescent accretion rates \citep[e.g.,][]{plotkin2017}. For a significantly inverted jet spectrum ($\alpha = 0.5$), this radio luminosity would only be a factor five lower. As neutron star HMXB jets are $\sim 3$ orders of magnitude radio fainter than those of black holes \citep{vandeneijnden2022} and the deepest limits on any accreting neutron star's radio emission in quiescence also lie far below $5\times10^{27}$ erg/s \citep{vandeneijnden2022b}, we deem this scenario unlikely. The mm non-detection of X1908+075 implies an upper limit on any $6$-GHz radio jet emission of $L_{\rm R} \leq 6\times10^{27}$ erg/s if $\alpha \geq 0$, as expected for a compact radio jet. This places X1908+075 amongst the sample of radio-non-detected neutron star SgXBs, with a limit below the typical radio luminosity of their radio-detected counterparts \citep{vandeneijnden2021}. 

For the remainder of this discussion, we will use the mm upper limit for X1908+075 as a limit on its thermal wind emission. Similarly, we will continue under the assumption that the mm detection of IGR J18410-0535 is dominated by its thermal stellar wind emission. We note that for the low state of SFXTs, the quasi-spherical settling accretion model proposes that accretion takes place via a hot extended plasma shell around the magnetosphere \citep[e.g.][]{shakura2012,shakura2014}, although other models for accretion cannot yet be ruled out \citep{grebenev2007,bozzo2008}. While this shell could potentially provide an additional site of low-frequency emission, detailed calculations of such emission are beyond the scope of this work.


\subsection{How to build upon this mm pilot study}
\label{sec:future}

\begin{figure}
\begin{center}
\includegraphics[width=\columnwidth]{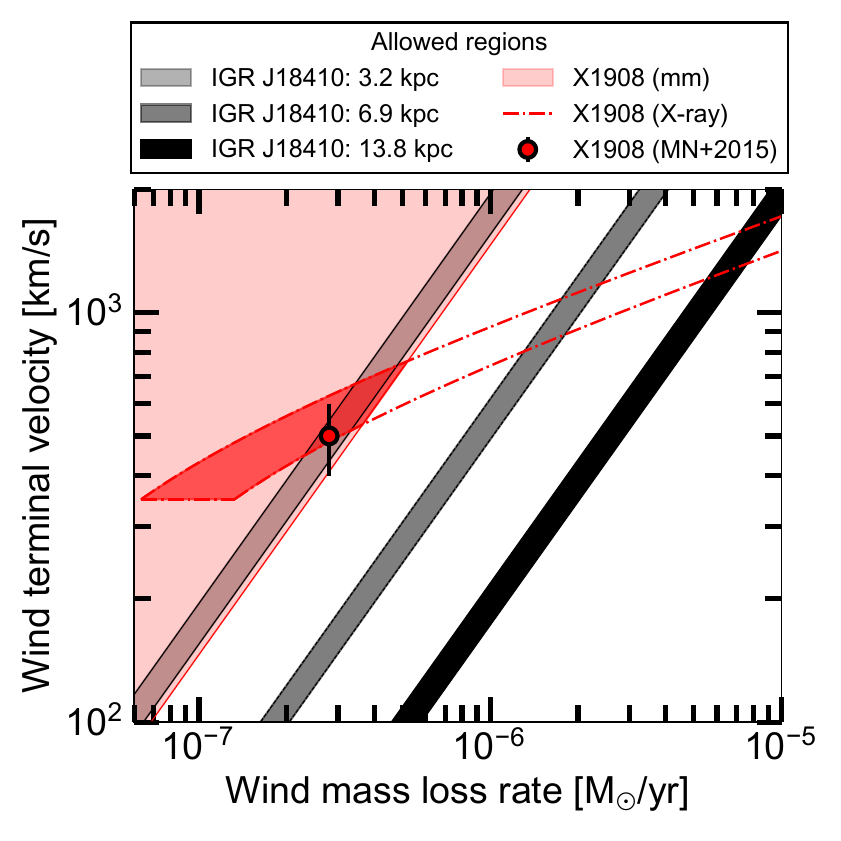}
\caption{Constraints on the stellar wind mass loss rate and terminal velocity of X1908+075 (X1908) and IGR J18410-0535 (IGR J18410). For the former, the constraints from the mm and X-ray bands are shown with the red region and lines, respectively, intersecting in the dark red region. The red circle shows the measurement for X1908+075 by \citet{martinez2015}, referred to as MN+2015. For IGR J18410-0535, the three grey/black shaded regions indicate the mm constraints for three distances reported in the literature (3.2, 6.9, and 13.8 kpc). See Section \ref{sec:future} for full details and a discussion of current caveats of this approach.}
\label{fig:wind}
\end{center}
\end{figure}

While our pilot study shows that neutron star HMXBs can be detected by current mm facilities, such a detection alone will not fully constrain the properties of its stellar wind. Crucially, the observed thermal continuum emission of a smooth stellar wind will depend on both its mass loss rate and terminal velocity (following e.g. \citealt{wright1975} and \citealt{panagia1975}):

\begin{equation}
    \frac{S_\nu}{7.26 \text{ mJy}} = \left(\frac{\nu}{10\text{ GHz}}\right)^{\frac{6}{10}} \left(\frac{\dot{M}_{\rm wind}}{10^{-6}\text{ }M_{\odot}\text{/yr}}\right)^{\frac{4}{3}} \left(\frac{\mu v_\infty}{100\text{ km/s}}\right)^{-\frac{4}{3}} \left(\frac{d}{\text{kpc}}\right)^{-2}
\label{eq:snu}
\end{equation}

\noindent where we assume that the wind has an electron temperature of $10^4$ K and a mean atomic weight of $\mu = 1$\footnote{The assumption on $T_e$ has little effect on our inferences, since $S_\nu \propto T_e^{0.1}$). Similarly, slight deviations in $\mu$ are masked by our larger systematic uncertainties in $v_\infty$.}. Evidently, the continuum emission only can therefore not uniquely constrain both mass loss rate and velocity; to highlight this challenge graphically, Figure \ref{fig:wind} shows the curves (IGR J18410-0535; light/dark grey and black for $3.2$, $6.9$, and $13.8$ kpc, respectively) and region (X1908+075; light red) in the $\dot{M}_{\rm wind}$--$v_\infty$ plane consistent with their NOEMA detection and upper limit, respectively. Not only can the mass loss rate and velocity not be uniquely determined based on mm information alone; the Figure furthermore reiterates that the distance uncertainty for IGR J18410-0535 implies that both targets of our NOEMA campaign currently overlap in the $\dot{M}_{\rm wind}$--$v_\infty$ plane -- only for larger distances, the SFXT curve is located in a region of this parameter space distinct from the SgXB. 

Going forward, however, other observational constraints may be folded into the analysis to separate both global properties. In this Section, we will therefore explore what complementary constraints could be applied for HMXBs in future studies and what limitations need to be overcome to successfully apply such a joint approach. For SgXBs, specifically, their accretion-driven X-ray behaviour may be folded in, as we can demonstrate for X1908+075: the same two wind parameters ($\dot{M}_{\rm wind}$, $v_\infty$), in combination with its orbital parameters, determine the wind capture rate in SgXBs and, in turn, the accretion rate and luminosity assuming standard Bondi-Hoyle-Lyttleton accretion. If the X-ray luminosity follows $L_X = \eta \dot{M}_{\rm acc} c^2$, where we will assume that $\eta = 0.1$, we can follow \citet{fkr2002} to write:

\begin{equation}
L_{X} = \dot{M}_{\rm wind} \frac{\eta c^2 G^2 M^2_{\rm NS}}{a^2 v_{\rm rel}^4} \text{ .}
\label{eq:lx}
\end{equation}

\noindent Here, $v_{\rm rel}$ is the relative velocity between the stellar wind and the neutron star, which we will approximate as $v_{\rm rel} \approx \sqrt{v^2_\infty + v^2_{\rm orbit}}$ and $v_{\rm orbit}$ is the orbital velocity of the neutron star. For our showcase X1908+075, \citet{martinez2015} reported a binary separation of $a \sin i = 1.43\times10^{12}$ cm and an eccentricity consistent with zero. Their derived inclination range between 46 and 58 degrees implies $a = (1.7-2.0)\times10^{12}$ cm. \citet{levine2004}, through independent methods, derived a consistent range in semi-major axis size between $a = (1.7-2.4)\times10^{12}$ cm. Here, we will adopt the latter (larger) range \citep[for comparison, the stellar radius is of the order $10^{12}$ cm;][]{martinez2015}. Its average orbital velocity is $v_{\rm rel} = 350 \pm 70$ km/s, where the error reflects the uncertainty on $i$ and therefore $a$. By adopting the X-ray luminosity derived in Section \ref{sec:results_xrays}, we can plot the edges of the X-ray allowed regions for X1908+075 in the $\dot{M}_{\rm wind}$--$v_\infty$ plane in Figure \ref{fig:wind} using the red dot-dashed lines (where we consider only the region where $v_\infty > v_{\rm orbit}$; see below). The overlap with the mm region is highlighted in darker red. 

This example of a combination of mm and X-ray constraints, assuming Bondi-Hoyle-Lyttleton accretion, can be compared with earlier, independent measurements: \citet{levine2004} and \citet{martinez2015} both derived estimates of the wind parameters of X1908+075. The former found a high mass loss rate of several times $10^{-6}$ $M_\odot$/yr (depending on the radial velocity profile) with a velocity around $800$ km/s, which is inconsistent with the constraints posed by the mm non-detection (although they fit with the X-ray constraints for this target). The latter work instead finds $\dot{M}_{\rm wind} \approx 2.8\times10^{-7}$ $M_\odot$/yr and a velocity $v_\infty = 500 \pm 100$ km/s, which lies within the region allowed by our mm and X-ray constraints. 

The application of this combined mm + X-ray approach in future campaigns comes with several caveats and issues to be assessed. Firstly, we have assumed that the wind has reached its terminal velocity at the orbital separation, which, for relatively small orbits and certain values of the wind's $\beta$-parameter, may not be the case. In that case, the $L_X$ from Equation \ref{eq:lx} is an underestimate, implying the corrected curve in Figure \ref{fig:wind} would shift upwards. Moreover, we only consider cases where the wind speed exceeds the orbital speed. If instead the two become comparable, the stellar wind will be significantly beamed toward the companion and a dense tidal stream forms \citep{blondin1991,elmellah2019a} which is not accounted for in the Bondi-Hoyle-Lyttleton computation of the mass accretion rate. In this case, the mass transfer proceeds through wind-RLOF \citep{mohamed2007} and the mass accretion rate can be enhanced according to semi-analytic \citep{abate2013} and numerical \citep{elmellah2019b} estimations. New (numerical) calculations of wind capture at low speeds and small orbits are therefore necessary to more accurately quantify the relation between wind properties and accretion rate for the entire relevant parameter space. We note that similar issues arise for SFXTs, which may be more complex to overcome. Equation \ref{eq:lx} for Bondi-Hoyle-Lyttleton accretion is likely not applicable to SFXTs, as it does not include the role of the magnetosphere and settling regimes as possible origins of their low persistent accretion rate \citep{shakura2012,Kretschmar2019}.

In our exploratory calculations for Figure \ref{fig:wind}, we have implicitly assumed a completely smooth wind, as Equation \ref{eq:snu} was derived under that assumption. While stellar wind are known to be inhomogeneous, which can lead to an enhancement of the wind’s thermal emission, we do not expect this to significantly impact a combined mm + X-ray approach. The level of enhancement depends on the level of clumping at the radius where a specific observed wavelength originates from; as clumping decreases as a function of radius \citep[e.g.][]{puls2006,martinez2017}, its effects are relatively minor at radio and mm bands. Modelling of clumped winds in the regime relevant for SFXTs and SgXBs suggests that the flux increase compared to a smooth wind is maximally of the order $50\%$ \citep[][]{blomme2011,daley2016}. If our measured NOEMA flux density (limit) is a factor $1.5$ higher than implied by the global wind parameters and Equation \ref{eq:snu}, it would at most correspond to an upward shift in the mm curves of Figure \ref{fig:wind} by 0.13 dex, which does not substantially affect the method. 

Our NOEMA pilot study, by design limited in several key aspects, has shown that neutron star HMXBs are detectable by current mm facilities and yielded the first low-frequency detection of a SFXT. Specifically, the mm detection and radio non-detection of the SFXT shows the advantage of the higher observing frequency in combination with the stellar wind's spectrum. In this Discussion, we have highlighted both the promise of mm studies of SFXTs and SgXBs to investigate their winds (and their possible differences) and the challenges to this approach, in particular when combining with X-ray constraints. On the observational side, to move beyond our pilot mm study, a significantly larger sample should be studied at mm sensitivities $\leq 10$ $\mu$Jy/bm. SgXB targets in follow up studies should, preferably, also have known orbits and X-ray behaviour, as well as coordinated cm-wavelength observations. Importantly, all targets should have well-constrained distances to allow for luminosity measurements limited by statistical uncertainties. In addition, new (numerical) calculations of the wind capture at small orbits and low speeds are necessary to improve the constraints offered by the X-ray band. Combined, such observational and theoretical developments may supply independent constraints to allow for a unique determination of their mass-loss rate and velocity. Current (sub-)mm observatories such as NOEMA and ALMA, as well as the planned next-generation VLA with its high sensitivity and spectral range up to 100 GHz \citep{selina2018}, will be vital in exploring these future opportunities. 

\section{Acknowledgements}
JvdE thanks J. Orkisz for guidance in NOEMA data reduction and analysis. For the purpose of open access, the author has applied a Creative Commons Attribution (CC-BY) licence to any Author Accepted Manuscript version arising from this submission. JvdE acknowledges a Warwick Astrophysics prize post-doctoral fellowship made possible thanks to a generous philanthropic donation, and was supported by a Lee Hysan Junior Research Fellowship awarded by St. Hilda’s College, Oxford, during part of this work. This publication has received funding from the European Union’s Horizon 2020 research and innovation programme under grant agreement No 101004719 (ORP). This work is based on observations carried out under project number W22BO with the IRAM NOEMA Interferometer. IRAM is supported by INSU/CNRS (France), MPG (Germany) and IGN (Spain). This research has made use of the MAXI data provided by RIKEN, JAXA, and the MAXI team. We thank Jamie Stevens and ATCA staff for the scheduling of these observations. ATCA is part of the Australia Telescope National Facility (\url{https://ror.org/05qajvd42}), which is funded by the Australian Government for operation as a National Facility managed by CSIRO. We acknowledge the Gomeroi people as the Traditional Owners of the ATCA observatory site. This research has made use of NASA's Astrophysics Data System Bibliographic Services.

\section*{Data Availability}

A GitHub reproduction repository will be made public upon acceptance at \url{https://github.com/jvandeneijnden/MM\_NS\_HMXB/}. The NOEMA observational data will become public three years after the final observation. See \url{https://iram-institute.org/science-portal/data-archive/} for instructions on access. The calibrated uv-tables (in \textsc{gildas} .UVT format) are available in the GitHub repository. Upon reasonable request, the authors may provide assistance in accessing uncalibrated data prior to public release.

\input{output.bbl}


\bsp	
\label{lastpage}
\end{document}